\%

\def\beq{\begin{equation}}
\def\eeq{\end{equation}}
\def\bea{\begin{eqnarray}}
\def\eea{\end{eqnarray}}
\def\bq{\begin{quote}}
\def\eq{\end{quote}}

\def\bseq{\begin{subequation}}  
\def\eseq{\end{subequation}}
\def\bsea{\begin{subeqnarray}}  
\def\esea{\end{subeqnarray}}

\def\gappeq{\mathrel{\rlap {\raise.5ex\hbox{$>$}}
{\lower.5ex\hbox{$\sim$}}}}

\def\lappeq{\mathrel{\rlap{\raise.5ex\hbox{$<$}}
{\lower.5ex\hbox{$\sim$}}}}

\def\bbz{fa Z \kern-8.9pt Z}

\documentstyle [12pt]{article}
\begin{document}
\thispagestyle{empty}
\vspace*{-2cm}
\begin{flushright}
{SLAC-PUB-6360}  \\
{CfPA 93-th-31} \\
{September 1993} \\
{T/E/AS} \\
\end{flushright}
\vspace{1cm}
\begin{center}
{\large Astrophysical Bounds} \\
\vspace{.2cm}
{\large on Milli-Charged Particles} \\
\vspace{.2cm}
{\large in Models with a Paraphoton}
\end{center}
\vspace{1cm}
\begin{center}
{Sacha Davidson\footnote{work supported by NSERC, and the National Science
          Foundation, grant AST 91-20005 } }\\
\vspace{.3cm}
{Center for Particle Astrophysics\\
University of California,  Berkeley, CA 94720}
\end{center}
\vspace{1cm}
\begin{center}
{Michael Peskin\footnote{work supported by the Dept of Energy, contract
DE-ACO3-76SF00515}}\\
\vspace{.3cm}
{Stanford Linear Accelerator Center\\ Stanford
University, Stanford, CA 94309 }
\end{center}
\hspace{3in}

\begin{abstract}
 The upper bound on the number of relativistic species present
at nucleosynthesis has been used to constrain particles with electric
charge $\epsilon e$ ($10^{-8} < \epsilon <1$). We correct
the bound previously calculated for milli-charged particles that interact
with a shadow photon.   We also discuss the additional constraints from
the properties of red giants and of Supernova 1987A.
\end{abstract}

\newpage

\paragraph{}

   A problem of continuing  interest in elementary particle physics
is   the possible existence of
particles with an electric charge  very small compared to the
electron charge.  We will refer to such particles as `milli-charged'
and will denote the small charge as $\epsilon \cdot e$.
  Traditionally, these particles have been considered as new
ingredients added to the standard model, reflecting
the mystery of the electric charge quantization observed in Nature.
In this context, many authors have pointed out
astrophysical constraints on the mass and charge of milli-charged
particles\cite{neu}--\cite{M+N}.  In addition,
a very persuasive argument against the existence of such particles
 is the fact that they would be forbidden
 in models with  a grand unification.

    However, in 1986,
 Holdom\cite{Bob1}  showed that, by adding a second, unobserved,
photon (the `paraphoton' or `shadow photon'), one could construct
grand unified models which contained milli-charged particles in a
natural way.   Holdom's scheme for milli-charged particles is
genuinely persuasive, and has stimulated new experimental
tests \cite{SLAC}.
It also requires a rethinking of the astrophysical constraints on
milli-charged particles\cite{DCB,Bob2,Bob3}.  In this paper, we will
improve on previous treatments of these constraints, correcting an
error in a previously claimed nucleosynthesis bound and discussing
possible additional bounds from stars and from Supernova 1987A.

  Our conclusions are presented in Fig. 1.  This figure is based on
Fig. 2 of ref. \cite{DCB}.  It includes the limits from direct
accelerator experiments \cite{oops},
 the Lamb shift, $\Omega < 1$, and other sources derived in ref.
\cite{DCB},  and                       the supernova bound from
\cite{M+R}, and adds the new limits from nucleosynthesis and
and from helium-burning stars described below.

Holdom  showed that particles with small electric
charge would appear naturally in grand unified models if the model
contained {\it two} unbroken $U(1)$ symmetries.
 Conventional grand unification
leads to one unbroken $U(1)$ symmetry; at low energies, this is the
gauge symmetry of electromagnetism.  Holdom suggested that it could
easily contain another unbroken $U(1)$ which lives completely outside
the standard model gauge group.   At the most
fundamental level, the model would contain as light fermions
ordinary quarks and leptons,
which are neutral under the shadow $U(1)$, and new shadow fermions
which are neutral under the ordinary $U(1)$.  However, any small
mixing of the two $U(1)$ gauge bosons, even if it is induced by
 loop diagrams involving particles at the grand unification
scale, will cause  observable milli-charges with $\epsilon$
proportional to the
mixing angle.  A natural size of this mixing angle is $\alpha/\pi \sim
10^{-3}$.  In the resulting field space, the original $U(1)$ directions
are not orthogonal by the amount $\epsilon$.  We will define the
conventional photon ($\gamma$)
to be that linear combination of the two gauge bosons
which
couples to ordinary quarks and leptons, and define the paraphoton
($\gamma'$)
to be the orthogonal combination of gauge fields.  Then the
conventional photon has couplings of size  $\epsilon$ to shadow matter,
but the shadow photon has zero coupling to conventional matter.  The
opposite convention is also possible, by taking a different choice of
basis in the field space, but it is less convenient.  The value of
the basic  $U(1)$    charge could well be different between the
 ordinary and shadow $U(1)$'s ($\alpha'$ could differ from
$\alpha$), but, for simplicity, we will ignore this distinction below.

  One of the strongest constraints on models of milli-charged particles
both with and without a paraphoton is the bound from primordial
nucleosynthesis.  If the mass of the milli-charged particle is
sufficiently small, this particle will provide extra light degrees of
freedom at the era of nucleosynthesis and will thus contradict the limit
usually quoted as the bound on the number of light neutrinos.
The principal uncertainty in this bound comes from the estimation of the
primordial mass fraction $Y_p$ of $^4$He. In a recent paper, Walker,
Steigman,  Schramm,
Olive, and Kang \cite{OW}  have argued for the relation
\beq
     N_\nu  =  2.00 \pm 0.15 + 83.3 (Y_p - 0.228) .
\eeq
and for the value
$Y_p = 0.23 \pm 0.01$.  However, the standard error given here
should be used
with care in citing confidence limits, since it is mainly systematic.
We believe it is correct, in citing limits on new particles, to
consider a scenario with $N_\nu = 4.2$  as acceptable, and we will
argue below in this spirit.  A much stronger conclusion would follow if
we took the error on $Y_p$ literally as the width of a Gaussian
distribution
and claimed $N_\nu < 3.3$ (95~\%
confidence); we will also discuss the bound in this case below.

A milli-charged fermion with a small mass counts as two neutrinos, to be
added to the three light neutrinos of the standard model.  Thus,
milli-charged particles with  $m_{\epsilon} \lappeq 1$ MeV can be
ruled out (for  $\epsilon \gappeq 10^{-8}$). This
bound applies to models both  with and without a
paraphoton.
However, it was incorrectly claimed in \cite{DCB} that the
nucleosynthesis bound in the model with a paraphoton was much stronger,
ruling out
$m\lappeq$ 200 MeV.  Since this is a region of parameter space in which
a direct experiment test of the model is possible\cite{SLAC}, we
should correct this conclusion.

The upper bound on the number of light neutrinos at nucleosynthesis is,
more correctly, a bound on the energy density at this epoch.  Since the
paraphoton is necessarily massless, it always contributes to this
energy density.  A paraphoton in thermal
equilibrium with ordinary photons has a contribution equal to 8/7 of
a neutrino, and so is excluded only by the strongest claimed bounds
on $N_\nu$.  However, the authors of \cite{DCB} argued that the
contribution of the paraphoton was considerably larger.
They assumed that, if milli-charged particles were heavier than 1 MeV,
these particles would annihilate before the era of nucleosynthesis and
transfer their entropy
 to the paraphotons.  This would raise the
paraphoton temperature with respect to the
photon temperature and increase the paraphoton energy density.
They concluded that
the milli-charged particles had to annihilate before the
QCD phase transition, where substantial entropy production increases the
temperature of ordinary photons.

  However, this is unneccessary.  We will now show that
the paraphotons remain in thermal equilibrium with the photons
as the milli-charged particles annihilate, so that $T_{\gamma'}$ will
never be significantly larger than  $T_{\gamma}$.

Photons can turn into paraphotons by
Compton scattering from a milli-charge
(or anti-millicharge)
  to
produce a paraphoton.  For a given photon, the
rate of this process is
given by
$\Gamma(\epsilon \gamma \rightarrow \epsilon \gamma') \simeq n_{\epsilon}
\sigma(\epsilon \gamma \rightarrow \epsilon \gamma')$, which we can
estimate for
$m_{\epsilon} < T$  by
\beq
\Gamma(\epsilon \gamma \rightarrow \epsilon \gamma') \simeq
 4 \left(\frac{m_{\epsilon} T}{2 \pi} \right)^{3/2}
e^{-m/T}\cdot
\frac{8 \pi \epsilon^2 \alpha^2}{3 m_\epsilon^2 } ,   \label{1}
\eeq
where we have used the low-energy limit of the Compton cross section
and assumed that $\alpha' = \alpha$. The rate for converting a
paraphoton to a photon is identical.
 To estimate quantitatively the
temperature at which the milli-charges have transferred their
entropy to the paraphotons, we define $T_\epsilon (m_{\epsilon})$
to be the the
temperature at which the equilibrium density of milli-charges
has fallen to 1/10 the number density of a relativistic species:
\beq
\left( \frac{m_{\epsilon} T_{\epsilon}}{2 \pi} \right)^{3/2}
e^{-m_{\epsilon}/T_{\epsilon}} = \frac{1}{10} \cdot
\frac{ 7 \zeta(3)}{8 \pi^2}
T_{\epsilon}^3~~.
\eeq
Then the paraphotons will equilibrate this entropy with ordinary
photons if the rate $\Gamma(\epsilon\gamma \rightarrow \epsilon
 \gamma')$, evaluated at $T_\epsilon$, is large compared to the
expansion rate of the universe at $T_\epsilon$:
\beq
\Gamma (\epsilon \gamma \rightarrow \epsilon \gamma')   >
H(T_\epsilon)
\simeq \frac{ 1.7 \sqrt{g_{eff}(T_{\epsilon})} T_{\epsilon}^2}{m_{pl}}
\eeq
where $H$ is the Hubble expansion rate, $m_{pl}$ is the Plank mass
and $g_{eff}(T)$  is the  total
 effective
number of relativistic degrees of freedom.
The inequality (3) is satisfied for $\epsilon >
10^{-8}$.

Thus, for $\epsilon> 10^{-8}$, nucleosynthesis gives
a lower limit to the milli-charge mass of 1 MeV, as before.  This revised
nucleosynthesis bound is plotted in Fig. 1.

If we accept the strong nucleosynthesis bound of ref. \cite{OW},
and neglect the possibility of a heavy (unstable?) tau neutrino
\cite{pre,Turner}, the
bound on the milli-charge mass becomes much stronger.  If one cannot
have more that 3.3 effective neutrinos at the era of nucleosynthesis,
the paraphotons cannot be in thermal equilibrium with the ordinary
 photons; rather, they must be cooler in such a way that
\beq
\frac{8}{7} T_{\gamma'}^4 = .3 T_{\gamma}^4~~.
\eeq
The photons must therefore be heated by annihilations after the
 paraphotons
decouple.    This implies \cite{DCB} that the photons and paraphotons
must be out of equilibrium at the temperature of the
QCD phase transition, $T_c \approx 200$ MeV.  The dominant equilibration
process is still Compton scattering, and so we can use the
estimates above, with $T_\epsilon$ replaced by $T_c$, to find the limit
\beq
m_{\epsilon} > 7 + .4 \ln \epsilon ~~{\rm GeV}~~.
\eeq
We plot this bound as a dotted line in figure 1.  We emphasize to the
reader that the difference between this limit and the  qualitatively
weaker limit above
corresponds to a 1 $\sigma$ shift in the $^4$He abundance constraint.

  Previous discussions of the astrophysical
bounds on milli-charged particles
made use of physical arguments which were independent of the
existence of the paraphoton.  In particular, when $\epsilon$ is small,
milli-charges made in the center of a helium-burning
star (or of Supernova  1987A)
can escape freely, adding substantially to the cooling
rate.  This leads to a bound for small masses and $\epsilon < 10^{-7}$.
However, in models with a paraphoton, there is another physical picture
which leads to constraints at larger values of $\epsilon$.

  The new bound arises when we consider the radiation of paraphotons
from the star.  It is not difficult to see that this radiation
 can be substantial:   Under circumstances that we will specify below,
the core of a star can contain milli-charges and  paraphotons in
thermal equilibrium with the ordinary matter.
  The cross section for paraphoton-millicharge
scattering is the full Compton cross section, without a factor
$\epsilon^2$.  Thus, the mean free path for paraphotons will be
much shorter than the size of the stellar core.  However, in outer
regions of the star, the temperature drops, the density of milli-charges
decreases, and the star becomes transparent to paraphotons.  Thus,
the star will have a photosphere for paraphotons  and will radiate
from this `paraphotosphere' approximately
like a black body. For milli-charged masses greater than  100 eV,
the paraphoton mean free path in the outer regions of the star
will be longer than that of the
photon, so the the `paraphotosphere' will occur at a smaller radius,
and thus a higher temperature, than the sphere from which ordinary
photons are radiated \cite{lowmass}.
Thus, the  stellar luminosity in paraphotons  should be  greater than
 the stellar luminosity in ordinary photons. However,
the sun is too young to have been losing energy
at twice the photon  luminosity,
and the observed lifetimes of helium-burning
stars also disfavour  such a large addition to the energy
production \cite{Raf}. We get a  better bound from the hotter
star, but a more dependable one from the sun. We will therefore
outline the calculation for a helium-burning star, but quote the
bound also for the analogous argument applied to the sun.

   Since milli-charges are produced from ordinary matter by
the pair production process $e^- N \rightarrow e^- N \epsilon
\bar \epsilon$, there is no difficulty in creating a thermal
population of milli-charges.  The weak link in the argument is the
bottleneck in converting ordinary photons to paraphotons if the
thermal density of milli-charges is small.   We can make a rough
estimate of this rate by assuming that paraphotons are only
created in a  helium core
of radius
$R_c \sim 10^9$ cm   in thermal equilibrium at the temperature
$T_c \sim 10$ keV.  The production rate of paraphotons in the
sphere by Compton scattering $\epsilon\gamma \rightarrow \epsilon
\gamma'$ is given by
\beq
\frac{4 \pi R_c^3}{3} n_{\gamma} n_{\epsilon} \sigma_{\epsilon
\gamma \rightarrow \epsilon \gamma'} \sim \alpha \alpha' \epsilon^2
R_c^3 T_c^4 \sqrt{\frac{T_c}{m_{\epsilon}}} e^{-m_{\epsilon}/T_c} .
\eeq
If we require that this be less than ${\cal L}_\gamma/T_s$, where
$ {\cal L}_\gamma$ is the luminosity of the sphere
in ordinary photons, and $T_s$
is the surface temperature, we find the limit
\beq
m_{\epsilon} \gappeq .4 + .02 \ln \epsilon ~ {\rm MeV~~~(He)} \label{He}
\eeq
for helium-burning stars, and
\beq
m_{\epsilon} \gappeq 40 + 2 \ln \epsilon ~ {\rm keV~~~(sun)}
\eeq
from the sun.
We have plotted (\ref{He}) as a thin line in Fig. 1.
Our new bound is obviously
very rough, but fortunately, it is unimportant, because it applies
only to a region of parameter space already ruled out by nucleosynthesis.

   A similar argument could in principle give a stronger bound from
the properties of Supernova 1987A.  However, the proto-neutron star
cools by emitting neutrinos, not photons. If $\epsilon$ is sufficiently
large that the milli-charged particles do not escape from the core,
the paraphoton mean free path is shorter than that of the neutrinos,
so the `paraphotosphere' would be at a larger radius (and lower
temperature) than the
`neutrinosphere', and the paraphoton luminosity would be lower than
that of neutrinos.   Thus, SN1987A gives no additional constraint in
models with a paraphoton beyond the limit found in ref. \cite{M+R} in
the model without a $\gamma'$.

   In this paper, we have presented new evaluations of the bounds on
milli-charged particles in models with a  paraphoton, as advocated by
Holdom.  Our main conclusion is that, with a conservative estimate of
the nucleosynthesis constraint, there is no astrophysical
restriction on the
existence of milli-charges with $m_\epsilon > 1$ MeV and $\epsilon <
10^{-2}$.  We look forward to new direct experiments which will
explore this interesting region.

  \bigskip
     We are grateful to John Jaros, Morris Swartz, and Willy Langenfeld
 for encouraging us to think about these issues, and also   to
  Fernando
Atrio-Barandela, Lance Dixon,
Pierre Salati and Martin White for informative discussions.

\newpage

\begin{figure}
\caption{Regions of mass-charge space ruled out for milli-charged
particles in the model with a paraphoton.  The bounds arise from
the following constraints: 1---accelerator experiments; 2---the Lamb
shift; 3---nucleosynthesis; 4---$\Omega < 1$; 5---plasmon decay in
red giants; 6---plasmon decay in white dwarfs; 7---dark matter searches;
8---Supernova 1987A; 9---$\gamma'$ emission by red giants.  The
bounds 1,2 and 4-7 are from
reference 6, 8 is from reference 5, and 3 and 9 are from this paper.}
\end{figure}


\begin{thebibliography}{222222}
\bibitem{oops} The values of $m_\epsilon$ excluded by direct
       searches  are quoted in Table II of \cite{DCB}, but the values
         given are in MeV, not GeV.  The limit is placed correctly,
         however, in Fig. 2 of \cite{DCB}.
\bibitem{lowmass}  It is unclear whether millicharges with
$m_\epsilon < 100$ eV would
be a serious drain on the energy of a burning star.  The
$\gamma'$---$\epsilon$ scattering cross-section increases as the
milli-charged particle gets lighter, so that below this point
 the paraphotosphere
might occur at a larger radius than the photosphere.  Fortunately,
this mass region is already excluded by the Lamb shift and by white dwarf
cooling \cite{DCB}.
\end{thebibliography}
\end{document}